\documentclass[debug,overfull]{epl}
\usepackage{epsfig}
\def\ltsim{\vbox {\hbox{\lower .8\baselineskip \hbox{$<$}} \break
                 \hbox{\lower 0.2\baselineskip \hbox{$\sim$}} } }

\title{Conductance of a molecular junction mediated by
unconventional metal-induced gap states}

\shorttitle{Conductance ...}
\author{R. Gutierrez\inst{1}, G. Fagas\inst{2},
K. Richter\inst{2}, F. Grossmann\inst{1} and R. Schmidt\inst{1}}
\institute{
\inst{1}
Institut f{\"u}r Theoretische Physik,
Technische Universit{\"a}t Dresden, D-01062 Dresden, Germany\\
\inst{2}
Institut f{\"u}r Theoretische Physik, Universit{\"a}t Regensburg,
D-93040 Regensburg, Germany
}

\pacs{73.63.-b}{Electronic transport in mesoscopic or nanoscale materials}
\pacs{85.65.+h}{Molecular electronic devices}
\pacs{73.22.-f}{Electronic structure of nanoscale materials; nanoscale contacts}

\begin{document}

\maketitle
\begin{abstract}
 The conductance of a molecular junction is commonly
 determined by either charge-transfer-doping, where alignment of
 the Fermi energy to the molecular levels is achieved, or tunnelling through
 the tails of molecular resonances within the highest-occupied
 and lowest-unoccupied molecular-orbital gap.
 Here, we present an alternative mechanism where electron transport is
 dominated by electrode surface states. They give rise to
 metallization of the molecular bridge and additional,
 pronounced conductance resonances allowing for
 substantial tailoring of its electronic properties via, e.g. 
 a gate voltage. This is demonstrated
 in a field-effect geometry
 of a fullerene-bridge between two metallic carbon nanotubes.
\end{abstract}

Molecular electronics has recently received increased
attention owing to the realization of active device components
based on single-molecule conductors at either low or room temperature.
These range from rectifiers, negative-differential-resistance
switches and electromechanical amplifiers to single-electron devices,
nanomechanical oscillators and field-effect transistors~\cite{Review}.
Such achievements have been complemented by advances in
the theory of electron transport across molecular junctions to include
both non-trivial quantum mechanical effects
at this scale~\cite{Datt95} (cf tunnelling and interference)
and the electronic structure of the components
involved~\cite{JG:CPL97,EK:PRB99,DTHRHK:PRL97,STDHK:96,LA:PRL00,NF:PRL99,RFCFSR:PRB02}.
Although some experimental aspects
remain controversial, detailed studies of current-voltage curves (I-Vs)
at either the semi-empirical or first-principles level
have accumulated evidence of what determines
the electronic response of an electrode--molecule--electrode setup.
The most important factors include the molecular electronic
structure~\cite{EK:PRB99},
band lineup and potential profile~\cite{DTHRHK:PRL97},
structural conformations \cite{STDHK:96,RFCFSR:PRB02},
charging effects \cite{LA:PRL00},
and the electron-phonon coupling in low-conduction
molecular wires~\cite{NF:PRL99}. Their respective action crucially depends  
on the realised interface.

In this Letter we discuss an important interfacial effect on the properties
of the molecular junction. Namely, we present an unconventional way to
metallization of a molecular-bridge, which owes to states
localized on the surface of the electrodes rather than to their
bulk Bloch states. They introduce prominent conductance
resonances within the HOMO-LUMO gap of the isolated molecule in close alignment to E$_F$,
in addition to the usual broadening of
the molecular levels by the metal-induced gap states (MIGS)~\cite{H:PR65}.
Their tuning by a gate voltage may be used to realize
a molecular switch in a field-effect geometry.

Surface resonant states of electrodes
are also important to scanning-tunnelling-microscope
(STM) images obtained by sharp tips~\cite{PHMYMMF:PRL98},
often employed in single-molecule transport
measurements~\cite{Review,JG:CPL97}.
However, such states are commonly missing from conventional
models of molecular junctions.
These usually involve the atomistic simulation of electrodes
with bulk or seamless properties, and the employment of idealised
jellium leads~\cite{JG:CPL97,EK:PRB99,DTHRHK:PRL97,STDHK:96,LA:PRL00,NF:PRL99}.

 \begin{figure} [t]
 \begin{center}
 \epsfig{figure=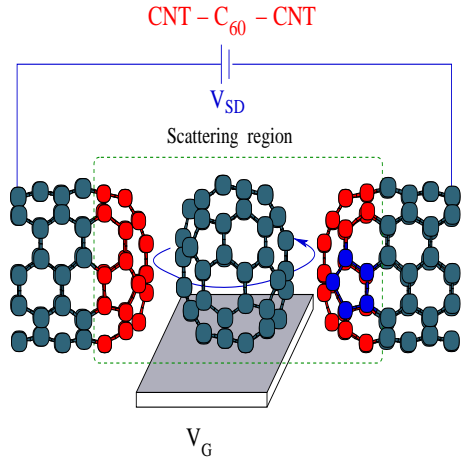, width=6cm, height=6cm}
 \caption{\label{fig1} All-carbon molecular junction composed of a \chem{C_{60}}
 between metallic CNTs. See text for details.}
 \end{center}
 \end{figure}
For definitiveness, we study the transport
properties of an all-carbon molecular device consisting of a carbon
nanotube (CNT)-\chem{C_{60}} hybrid system with {\it closed} end CNT surfaces.
The implications of our
results and experimental realizations are discussed later on.
Electron transport across the  molecular junction is studied
in the Landauer formalism~\cite{Datt95} by using numerical Green
function techniques
combined with a tractable electronic-structure tight-binding approach
parametrized by density functional theory (DFT)~\cite{basis}.
The two-terminal conductance as calculated by $G =({2 e^2}/{h})  T(E_{\rm F})$
is a function of the energy of injected electrons from the electrodes.
$T(E_{\rm F})$ is the transmission function, which
within the Landauer picture relates the conductance of the system to
an independent-electron scattering problem. The factor two implies spin-degeneracy.

We present the CNT--fullerene device-setup in Fig.~\ \ref{fig1}. We consider a
\chem{C_{60}} molecule between two semi-infinite metallic CNTs.
The diameters of the considered armchair-$(5,5)$ CNTs
and the fullerene are comparable ($\sim 0.7$nm).
In particular, a $(5,5)$ CNT is terminated by a cap with the shape
of half a \chem{C_{60}} molecule
positioned so that the nanotube axis crosses vertically
a single pentagon at its centre.
The CNTs are aligned mirror-symmetrically to a single-axis.
The centre of the fullerene is placed at the
mid-point between the two pentagons at the cap-edges. Although both
the \chem{C_{60}} and the capped CNTs have appreciable stability,
we perform additional structural optimization in the depicted scattering
region of Fig.~\ \ref{fig1} to include
any conformational changes due to the proximity of the
two subsystems. Mainly, the \chem{C_{60}} is compressed
along the transport direction taking an ellipsoidal shape.
For a device placed on a substrate the optimization details
depend on the substrate choice and deserve a systematic
study. Our main conclusions remain unaffected,
if a rather passive substrate is assumed. For example,
the surface van der Waals forces on CNTs deposited
on the H-passivated Si(100) modifies their shape and electronic
properties only if the diameter exceeds $\sim 3$nm \cite{HWA:PRB98}.

 \begin{figure} [t]
 \begin{center}
 \epsfig{figure=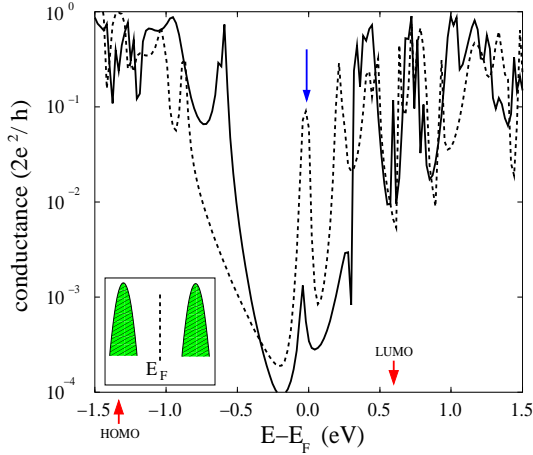, width=7cm, height=6cm}
 \caption{\label{fig2}Typical conductance spectra:
  the fullerene is positioned with either a pentagon (dashed curve) or
  a hexagon (solid curve) facing the pentagon of the left CNT cap.
  The HOMO and LUMO levels of the isolated \chem{C_{60}} are indicated by arrows.
  The marked resonance at E$_F$ is the main subject of this Letter.
  Inset shows schematically the conductance spectrum when the Fermi energy
  lies within the gap between the broadened levels of the HOMO and LUMO.}
  \end{center}
  \end{figure}
We emphasize three features of our system:
(i) it consists purely of carbon atoms, (ii) the lateral dimension
of the electrodes is at the nanoscale, and most
importantly, (iii) the surface of the electrodes, i.e, the caps,
possess unambiguous localised states with energies near
the equilibrium Fermi energy.
The latter have been predicted in Ref.~\cite{TT:PRB95}
and after experimental observation~\cite{PRL:CRACBVC97}
thoroughly studied for their field-emission properties.
All the above points have important implications for
possible applications. In anticipation of our results,
the fulfillment of (iii) for an electrode is a necessary condition
for the presence of unconventional MIGS in the molecular bridge.

It is instructive to first examine
typical spectra of the Landauer conductance $G$
for two possible orientations of \chem{C_{60}} with respect to the CNTs.
In Fig.~\ \ref{fig2}, the sensitivity of
electron transport on the geometry of the CNT electrodes--\chem{C_{60}}
interface is shown.
Similar to the results of our previous study~\cite{RFCFSR:PRB02}
of the conductance as a function of the relative orientation
in an {\it open} end CNTs-\chem{C_{60}} hybrid,
conductance variations up to two orders of magnitude are found.
The strong dependence on orientation derives from the interplay
between direct change of the magnitude of tight-binding coupling
parameters~\cite{KB:PRB02} and
more subtle effects reflecting the exact atomic-contact geometry~\cite{FCR:PRB01}. The latter include interference at the atomic level and
are typically enhanced at molecular junctions with
mesoscopic electrodes (see point (ii) above).

Detailed underlying effects are better demonstrated
in simpler setups by assuming only one molecular site coupled
to each electrode \cite{FCR:PRB01}. However, a hint is given by writing

\begin{equation}
\label{eq1}
T(E)={\rm Tr}[{\bf \Gamma}_L {\bf G}^r {\bf \Gamma}_R {\bf G}^a].
\end{equation}
${\bf G}^{r(a)}$ is the retarded(advanced) molecular
Green function including a self-energy interaction
${\bf \Sigma}_{L,R}$ (left, L, and right,R, lead)
with the electrodes.
The matrices ${\bf \Gamma}_{L,R}$ are defined via

\begin{equation}
\label{eq2}
{\bf \Gamma}_{L,R}=i[{\bf \Sigma}_{L,R}-{\bf \Sigma}_{L,R}^\dagger]
\end{equation}
\begin{equation}
\label{eq3}
{\bf \Sigma}_{L,R}={\bf J}_{L,R}^\dagger {\bf G}^r_{L,R} {\bf J}_{L,R}
\end{equation}
with ${\bf J}_{L,R}$ the coupling matrix of \chem{C_{60}} to the CNTs
and ${\bf G}^r_{L,R}$ the retarted Green function of the CNTs.
In our basis, ${\bf J}_{L,R}$ contains the DFT parametrised tight-binding
matrix elements. Rows of ${\bf J}_{L,R}$ couple molecular
sites to more than one CNT-site bringing in interference terms.

However, and in strong contrast to Ref.~\cite{RFCFSR:PRB02}, 
our central result relates to the marked peak in both curves of Fig.~\ \ref{fig2}. 
Whereas the peak-height evidently fluctuates strongly,
its relatively large width ($\sim 0.1$eV) compared to room temperature and
its adjacent position to the Fermi energy prevail for a large number of possible
orientations that we have investigated. This suggests
an effective means to control the electronic properties of the device
(see below). We will now analyse the origin of these states.

 \begin{figure} [t]
 \begin{center}
 \epsfig{figure=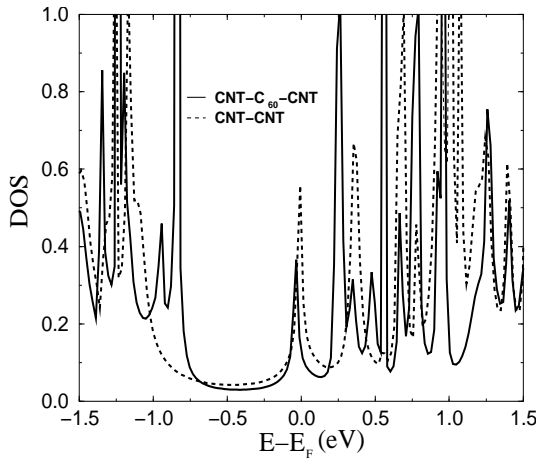, width=7cm, height=6cm}
 \caption{\label{fig3} Density of states per site (a.u.) of the indicated
 molecular junctions. The peak in the vicinity of E$_F$ is attributed
 to resonant cap states.}
 \end{center}
 \end{figure}
If no significant charge transfer occurs between a molecular bridge
and the electrodes, which could pin E$_F$ to one of the molecular
resonances, then electron transport proceeds by tunnelling through
the HOMO-LUMO gap as indicated in the inset of Fig.~\ \ref{fig2}.
Such band lineup is typical for metal--semiconductor interfaces leading
to a Schottky barrier. Otherwise, if charge transfer shifts the 
position of the molecular resonances, the conductance around E$_F$
would look similar to Fig.~\ \ref{fig2}, which would point to
resonant tunnelling via a molecular orbital.
This would however lead to a misinterpretation of our results. 
As inferred by the pure-carbon structure (point (i) above), we can safely 
exclude the possibility of large charge-transfer-doping at 
equilibrium conditions in the present set-up. This is further supplemented 
by Mulliken population analysis showing a small accumulation of charge
on \chem{C_{60}} ($\sim0.01e$).
Note aditionally that by
the position of the HOMO and LUMO of the isolated \chem{C_{60}}
alone, we cannot conclude to which molecular resonance, if any, the peak at E$_F$
corresponds. After contact to the electrodes
the broadened molecular levels undergo positional shifts.
This effect is enhanced by the multifold
degeneracy of the HOMO (5-fold) and LUMO (3-fold).
For the above reasons we argue that
the observed peak is a manifestation of resonant states of the CNT caps,
which lie within the HOMO-LUMO gap.

To further test the nature of the observed conductance resonance in the
vicinity of E$_f$ we have examined the local DOS.
It is interesting to note that in a simple setup the transmission function
reduces to the familiar formula
$T(E)\propto \Gamma_L \Gamma_R \nu_L \nu_R |G^r_{1N}|^2$, where
$\nu_{L,R}$ is the local DOS at a cap-site~\cite{FCR:PRB01}.
In Fig.~\ \ref{fig3}, we plot the local DOS per site,
$\nu(E)=-\frac{1}{\pi N} \Im Tr [{\bf G}(E){\bf S}]$,
where N is the number of atomic sites over which the trace is taken.
The overlap matrix ${\bf S}$ reflects the non-orthogonality of our basis.
For the $\nu(E)$ depicted in Fig.~\ \ref{fig1} the average is
taken over the scattering region and one CNT unit cell on
each side for two possible
junctions. One setup is that of Fig.~\ \ref{fig1}. In the other, the
\chem{C_{60}} molecular bridge has been removed. The appearance of the peak
around the Fermi energy in both curves suggests that the conductance resonance
in discussion corresponds to a cap state and not to a molecular orbital.
We note that the bulk DOS of a metallic CNT shows a plateau in this spectral
window.

 \begin{figure} [t]
 \begin{center}
 \epsfig{figure=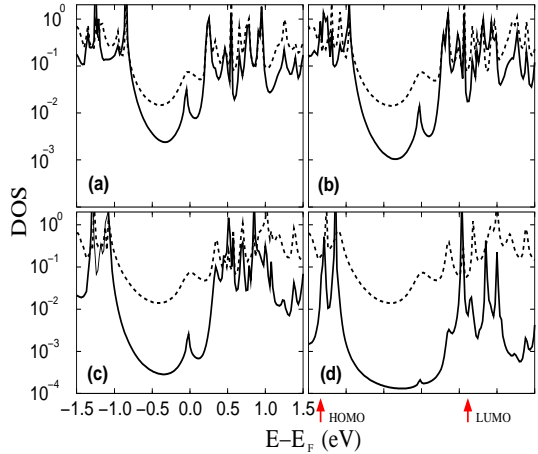, width=7cm, height=6cm}
 \caption{\label{fig4} Evolution of the density of states (a.u.) of the CNT cap
 and the \chem{C_{60}} (dashed and full curve, respectively)
 for decreasing mutual interaction ((a) to (d)).}
 \end{center}
 \end{figure}
It is evident that even when the conductance is dominated
by the cap states, electron transport should be
supported by the molecular bridge. Namely, the wavefunction of the cap resonances
must have noticeable amplitude on the molecule via evanescent
penetration, in a similar fashion to through-bond tunnelling of Bloch
states of the electrodes. Hence, the DOS of the
fullerene should not only be finite but also show a similar peak.
That is indeed the case as shown in Fig.~\ \ref{fig4}.
When increasing the distance between the electrodes in
the setup of Fig.~\ \ref{fig1} the DOS of the \chem{C_{60}}
resembles that of its slightly perturbed spectrum.
Both the small weight of the \chem{C_{60}} DOS-peak at E$_F$
and its gradual reduction without any appreciable shift point again
to the fact that it cannot be associated to either a HOMO or LUMO.

Having established the origin of the conductance resonance
adjacent to E$_F$ and the role of the \chem{C_{60}} molecular
bridge as a through-bond mediator, one may utilise the latter
to efficiently control the conductance of such a molecular device.
Rotation of the \chem{C_{60}} as implied by Fig.~\ \ref{fig2}, and previously
suggested~\cite{RFCFSR:PRB02}, or other mechanical manipulation such as
the demonstrated compression
by an STM tip may be applied~\cite{JG:CPL97}.
However, the resonant position
with respect to E$_F$ makes the application of a gate potential
V$_g$ readily accessible. This probe may be provided by a third
CNT~\cite{WM:APL01} or a gated STM tip~\cite{GCK:APL00}.

\begin{figure} [t]
 \begin{center}
 \epsfig{figure=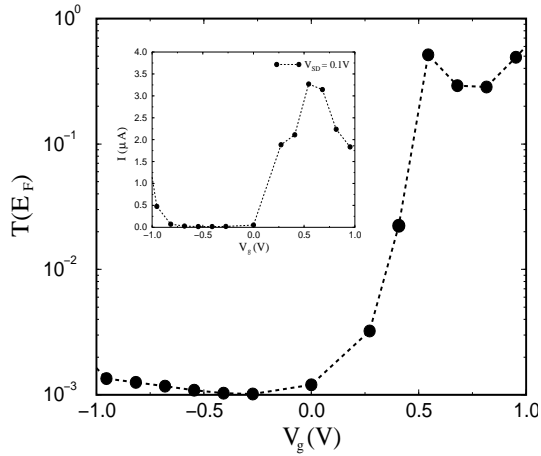, width=7cm, height=6cm}
 \caption{\label{fig5} Transmission function at E$_F$ as a function of the gate 
voltage V$_g$.
 Inset shows the current for a  fixed
 source-drain bias V$_{SD}$.}
 \end{center}
 \end{figure}

An external electric field modifies the charge-density inside
the molecular region, and, hence, the electronic coupling across
the molecular junction. We apply a crude approximation just to validate
its effect on transport properties by rigidly shifting the \chem{C_{60}} spectrum
by the gate voltage. In Fig.~\ \ref{fig5}, both the transmission function at
E$_F$ and the current via
$I=(2e/h)\int_{E_f-eV_{SD}/2}^{E_f+eV_{SD}/2}T(E) dE$
for source-drain voltage $V_{SD}=0.1$V are plotted
as a function of V$_g$. Manifestation of effective molecular switching is
apparent.

Because of strong coupling between the \chem{C_{60}} and the electrodes,
as indicated by the finite molecular spectrum in Fig.~\ \ref{fig4}
at all energies, no Coulomb blockade-related effects are expected.
Electron interactions are registered only through the exact
charging of the molecule via the electrostatic potential profile
determined in a self-consistent manner. As we mentioned, in our
system this defines nothing more than the details of electronic response
when a gate-voltage is applied, also at moderate
source-drain voltages ($\sim 0.1$eV). 

The suggested design not only serves as a definite example of unconventional
molecular metallisation, but it is also of interest to the search for
new-type all-carbon molecular devices. For example,
much success has been made in realizing either single-fullerene or CNT devices.
But unlike Fig.~\ \ref{fig1} in most experiments
contact is made with common bulk metals. In order to take full advantage
of molecular electronics prospects both the active region and the wiring
elements should be at the molecular scale. At present this role is reserved for CNTs,
also in our setup. Experiments already include I-V measurements
of a carbon nanotube ring and a DNA molecule in an STM setup~\cite{WM:APL01}. In those,
CNTs are used for STM manipulation and simultaneously provide
source/drain reservoirs and a gate terminal. More recently, interconnected
CNTs by molecular linkers have been reported~\cite{CDDR:APL02}.

In conclusion, we have employed a novel molecular device
to point out the existence of an alternative conduction mechanism in 
molecular junctions via unconventional metal-induced gap states.
The latter owe exclusively to resonant states of the
electrode contacting-interface and contrast the adopted view
that electron transport across molecular junctions is simply dominated by
hybridisation of the molecular orbitals. Such metallization is generic
to the appearance of surface resonant states within the HOMO-LUMO gap. Switching 
behaviour under a gate voltage was demonstrated in this device.
Moreover, anchoring groups of the molecular bridge may possess such states.
One should further address their exact properties in the environment of a
molecular junction together with their fingerprints on transport. 
Thus, we hope that the possibility
to effectively tailor the electronic response of such setups
will stimulate experiments and closer ties of molecular electronics
to surface science.

\acknowledgments
This study was supported by the Deutsche Forschungs\-gemeinschaft
through the Forschergruppe ``Nanostrukturierte Funktionselemente
in makroskopischen Systemen''.
GF is grateful to the Alexander von Humboldt foundation.


\begin{thebibliography}{99}

\bibitem{Review}
Aviram A., Ratner M. and Mujica V. (Editors),
{\it Molecular Electronics II}, Ann. N.Y. Acad. Sci. {\bf 960} (2002);
Joachim C., Gimzewski J.~K. and Aviram A., Nature (London), {\bf 408}
(2000) 541; for a recent account see: H{\"a}nggi P., Ratner M.
and Yaliraki S.~N. (Editors), Special Issue of Chem. Phys. {\bf 281} (2002).

\bibitem{Datt95}
Datta S., {\it Electronic Transport in Mesoscopic Systems} (Cambridge
University Press, Cambridge, 1995).

\bibitem{JG:CPL97}
Joachim C., Gimzewski J.~K. and Tang H., Phys. Rev. B, {\bf 58} (1998) 16407.

\bibitem{EK:PRB99}
Emberly E.~G. and Kirczenow G., Phys. Rev. B , {\bf 60} (1999) 6028;
Di Ventra M., Pantelides S.~T. and Lang N.~D., Phys. Rev. Lett., {\bf 84} (2000) 979.

\bibitem{DTHRHK:PRL97}
Datta S., Weidong T., Hong S. {\it et al} Phys. Rev. Lett., {\bf 79} (1997) 2530;
Yaliraki S.~N., Roitberg A.~E., Gonzalez C. {\it et al}, J. Chem. Phys., {\bf 111}
(1999) 6997; Xue Y., Datta S., and Ratner M.~A., J. Chem. Phys., {\bf 115} (2001) 4292;
Nitzan A., Galperin M., Ingold G.-L. and Grabert H., to appear in J. Chem. Phys..

\bibitem{STDHK:96}
Samanta M.~P., Tian W., Datta S. {\it et al}, Phys. Rev. B, {\bf 53} (1996) R7626.

\bibitem{LA:PRL00}
Lang N.~D. and Avouris Ph., Phys. Rev. Lett., {\bf 84} (2000) 358;
Taylor J., Guo H. and Wang J., Phys. Rev. B, {\bf 63} (2001) 121104(R);
Lehmann J., Ingold G.-L. and H{\" a}nggi P., Chem. Phys. {\bf 281}, (2002) 199.

\bibitem{NF:PRL99}
Ness H. and Fisher A.~J., Phys. Rev. Lett., {\bf 83} (1999) 452.

\bibitem{RFCFSR:PRB02}
Gutierrez R., Fagas G., Cuniberti G.
{\it et al}, Phys. Rev. B, {\bf 65} (2002) 113410.

\bibitem{H:PR65}
For MIGS in bulk semiconductors see:
Heine V., Phys. Rev., {\bf 138} (1965) A1689;
Louie S.~G. and Cohen M., Phys. Rev. B, {\bf 13} (1976) 2461;
for MIGS in nanowires see: Landman U., Barnett R.~N., Scherbakov A.~G.,
Avouris Ph., Phys. Rev. Lett., {\bf 85} (2000) 1958;
Tomfohr J.~K. and Sankey O.~F., Phys. Rev. B, {\bf 65} (2002) 245105.

\bibitem{PHMYMMF:PRL98}
V{\'a}zquez de Parga A.~L., Hern{\'a}n O.~S., Miranda R. {\it et al},
Phys. Rev. Lett., {\bf 80} (1998) 357, and {\it refs therein}.

\bibitem{basis}
A DFT parametrization of nonorthogonal linear
combination of (valence) atomic orbitals is used, see e.g.
Gutierrez R., Grossmann F., Knospe O. and Schmidt R.,
Phys. Rev. A, {\bf  64} (2001) 013202, and {\it refs therein}.

\bibitem{HWA:PRB98}
Hertel T., Walkup R.~E., and Avouris Ph., Phys. Rev. B
{\bf 58}, (1998) 13870.

\bibitem{TT:PRB95}
Tamura R. and Tsukada M., Phys. Rev. B, {\bf 52} (1995) 6015.

\bibitem{PRL:CRACBVC97}
Carroll D.~L., Redlich P., Ajayan P.~M.
{\it et al}, Phys. Rev. Lett., {\bf 78} (1997) 2811.

\bibitem{KB:PRB02}
Kornilovitch P.~E. and Bratkovsky A.~M., Phys. Rev. B, {\bf 64} (2001) 195413.

\bibitem{FCR:PRB01}
Fagas G., Cuniberti G. and Richter K., Phys. Rev. B, {\bf 63} (2001) 045416;
Cuniberti G., Fagas G. and Richter K., Chem. Phys., {\bf 281} (2002) 465;
Latg{\'e} A., Marcucci D.~C. and Tovar Costa M.~V., Physica E, {\bf 13} (2002) 1264.


\bibitem{WM:APL01}
Watanabe H., Manabe C., Shigematsu T. {\it et al}, Appl. Phys. Lett., {\bf 78}
(2001) 2928; {\bf 79} (2001) 2462.

\bibitem{GCK:APL00}
Gurevich L., Canali L. and Kouwenhoven L.~P., Appl. Phys. Lett., {\bf 76} (2000) 384.

\bibitem{CDDR:APL02}
Chiu P.~W., Duesberg G.~S., Dettlaff-Weglikowska U. and Roth S.,
Appl. Phys. Lett., {\bf 80} (2002) 3811;
there, interesting geometric effects arise \cite{FCR:PRB01}.

\end{thebibliography}
\end{document}